# Appropriate Observables for Investigating Narrow Resonances in Kaon Photoproduction off a Proton[*]

T. Mart

*Departemen Fisika, FMIPA, Universitas Indonesia, Depok 16424, Indonesia*
*tmart@fisika.ui.ac.id*

**Abstract.** The existence of non-strange partner of pentaquark, the $J^p = \frac{1}{2}^+$ narrow resonance, has been investigated by utilizing kaon photoproduction off a proton. It is found that the corresponding mass is 1650 MeV and the appropriate observables for investigating this resonance are the recoiled hyperon polarization, the beam-recoil double polarization $C_x$, and differential cross section at backward angles. Future kaon photoproduction experiments at low energies should focus on these observables.

**Keywords:** Kaon photoproduction, narrow resonance, antidecuplet.
**PACS:** 13.60.Le, 13.30.Eg, 14.20.Gk, 25.20.Lj

## INTRODUCTION

The nonstrange partner of pentaquark predicted by the chiral quark soliton model (χQSM), which is originally called the $N(1710)$ resonance (see Fig.1), has a strong decay width to the $\eta N$ channel and angular momentum state $P_{11}$ [1]. However, it has also sizeable decay widths to the $\pi N$ and $K\Lambda$ channels, which therefore provide at least three processes for investigating its existence.

The possibility of its decay to the $\pi N$ channel has been immediately investigated, after the LEPS collaboration reported the observation of pentaquark [2], by using a modified partial wave analysis (modified PWA), since a standard PWA can miss such a resonance [3]. In the $\eta N$ photoproduction off a free neutron a substantial enhancement of cross section at $W = 1670$ MeV is confirmed by experiments [4], which could be interpreted theoretically as a direct evidence of this resonance, although different interpretations such as interference effects from well established nucleon resonances are also possible [5].

It is surprising that before our previous recent papers [6,7] there had been no investigation on the existence of this resonance in the $K\Lambda$ channel. This is presumably due to the higher kaon mass, which makes this reaction channel more difficult to study, both theoretically and experimentally.

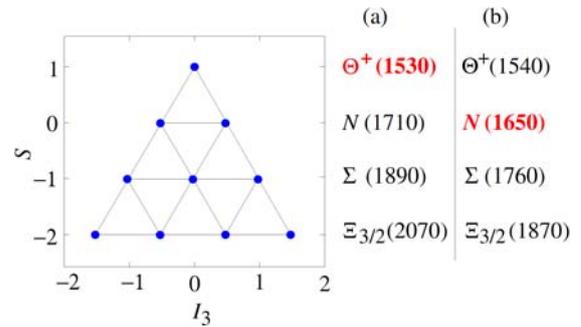

**FIGURE 1.** Masses of the antidecuplet member as (a) suggested by Ref. [1] and (b) obtained in our work with the mass splitting of 110 MeV [7].

In our recent papers [6,7] we have reported the result of our investigation on this resonance by utilizing the $K\Lambda$ photoproduction off a proton. Since the predicted resonance mass is very close to the reaction threshold, we found that a simple model that can accurately describe the process at low energies is more useful than a model that fits all experimental data with the energy up to 2.5 GeV, but tends to overlook small structures indicated by experimental data near the threshold region. By using such a simple model we have investigated the existence of this narrow resonance in kaon photoproduction. It is found that the corresponding mass is 1650 MeV. Starting from the mass of this resonance we predict masses of other antidecuplet members as shown in Fig.1.

In this paper we briefly review the finding in our previous papers and present the most promising observables to observe this narrow resonance in kaon photoproduction. For more detailed formalism and discussion of the isobar models used in this paper we refer the reader to our previous reports [6,7].

## ELEMENTARY PHOTOPRODUCTION MODEL

In this study the photoproduction process
$$\gamma + p \rightarrow K^+ + \Lambda$$
is described by an isobar model for low energy region as discussed in details in our previous paper [8]. The background amplitudes of this model is constructed from the standard $s$-, $u$-, and $t$-channel Born terms along with the $K^*(892)$ and $K_1(1270)$ $t$-channel vector mesons. In addition, we have also included the $S_{01}(1800)$ hyperon resonance in order to improve the agreement with experimental data [8]. The Feynman diagrams for this background are shown in Fig.2. Since the energy of interest is limited up to 1730 MeV, only six nucleon resonances might contribute to the process, i.e. the $S_{11}(1650)$, $D_{15}(1675)$, $F_{15}(1680)$, $D_{13}(1700)$, $P_{11}(1710)$, and $P_{13}(1720)$. All these resonances are considered in the model by using the resonant electric multipoles in the Breit-Wigner form. Nevertheless, as reported in our previous paper [6], contribution from the $S_{11}(1650)$ state is proven to be dominant.

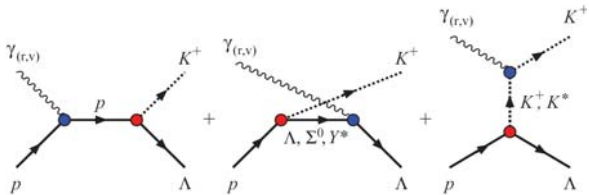

**FIGURE 2.** Feynman diagrams for the background amplitudes.

The leading coupling constants $g_{K\Lambda N}$ and $g_{K\Sigma N}$ are fixed by the SU(3) symmetry, whereas other hadronic ones are extracted from the available experimental data. Coupling constants of the nucleon resonances are taken from the Particle Data Book [9], whereas the phase angles are considered as free parameters. It is also important to emphasize here that hadronic form factors are not required in this model, since the energy considered is sufficiently low and therefore the intermediate states in the process propagate close to their on-mass-shell. The exclusion of this form factor substantially simplify the reaction amplitudes and, hence, remove some uncertainties in the model.

Result of the fitting to experimental data is shown in Fig.3, where we compare differential cross sections calculated from two different isobar models, obtained merely from different strategies in limiting the free parameters in the models, with Kaon-Maid [10] and experimental measurement [11-13]. In the fitting database, there exist 704 data points within the energy range of interest, i.e. from threshold up to $W = 1730$ MeV. By limiting the values of resonance parameters in the fit to vary within 10% of their original PDG values, the total $\chi^2$ can be reduced to 859, i.e. $\chi^2/N = 1.22$ (Model 1). Smaller $\chi^2/N$ (i.e. 1.00) would be obtained if we removed this limit (Model 2). As shown in Fig.3, both models provide a significant improvement to the Kaon-Maid [10].

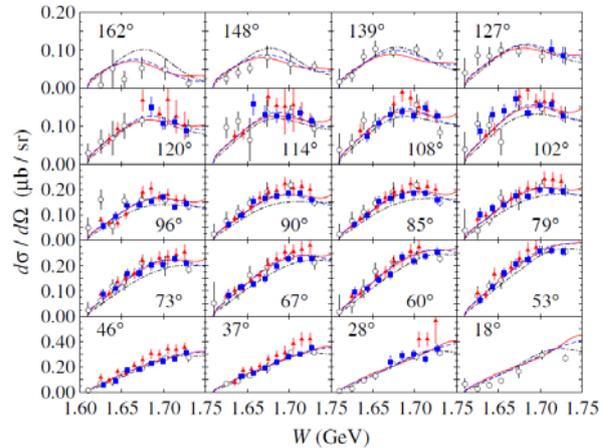

**FIGURE 3.** Comparison between differential cross sections calculated from Model 1 (solid lines), Model 2 (dashed lines) and Kaon-Maid [10] (dash-dotted lines) with experimental data from the SAPHIR (open circles) [11], CLAS2006 (solid squares) [12] and CLAS2010 (solid triangles) [13] collaborations. The corresponding kaon scattering angle is shown in each panel [6].

## RESULTS AND DISCUSSION

Having obtained reliable models for elementary process of kaon photoproduction, we insert a narrow nucleon resonance in the model and fit its free parameters, i.e. the helicity photon coupling, kaon branching ratio, as well as the phase angle, by fixing its mass and total width. This procedure is repeated for different values of resonance mass and width. In order to obtain the position of resonance mass and width with the smallest $\chi^2$, more than 1000 fits have been performed. In Fig. 4, the change of $\chi^2$ after the inclusion of the $P_{11}(1650)$ state in both isobar models is exhibited, where we can clearly see that the minimum at $m_{N^*} = 1650$ MeV seems to be model independent. This is in contrast to other minima at 1700 and 1720 MeV. The significant improvement of $\chi^2$ shown in Fig.4. indicates that a narrow resonance with a certain total width is required in order to explain the experimental data.

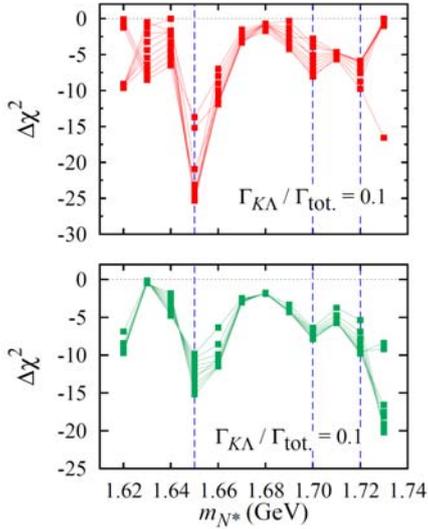

**FIGURE 4.** Change of the $\chi^2$ in the fit of Model 1 (upper panel) and Model 2 (lower panel) due to the inclusion of the $P_{11}$ narrow resonance as a function of its mass.

The possibility that the indication of the narrow resonance in kaon photoproduction originates from another state, i.e. the $P_{13}(1650)$ state, has been also investigated [6]. In this case, it is found that the minimum at 1650 MeV becomes weaker and another minimum at 1680 MeV appears. Thus, the structure appearing at 1650 MeV in the hyperon polarization data presumably does not originate from this state.

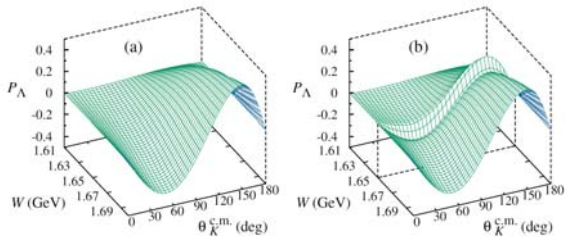

**FIGURE 5.** Recoiled hyperon polarization $P_\Lambda$ as functions of total c.m. energy $W$ and kaon scattering angle $\theta$ obtained from calculations (a) without and (b) with the narrow resonances $P_{11}(1650)$.

As reported in our previous paper [6], the minimum shown in Fig.4 originates from the recoiled $\Lambda$ polarization data. In Fig.5 we plot the clear effect of this narrow resonance, where we compare the calculated polarization with and without this resonance. Although this observable indicates a clear structure at 1650 MeV, new measurements with higher statistics are required in order to precisely constrain the mass, width, and helicity photon couplings of this resonance.

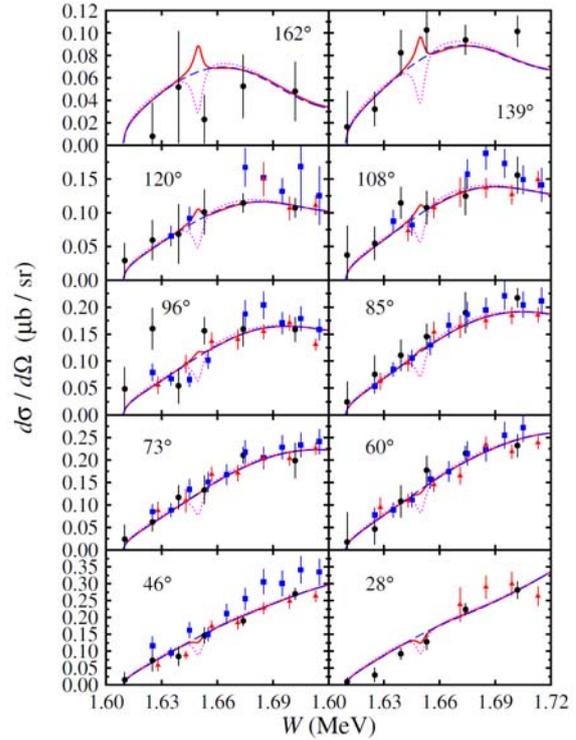

**FIGURE 6.** Energy distributions of the differential cross section obtained from the calculation without a narrow resonance (dashed lines), with the $P_{11}(1650)$ (solid lines), and with the $S_{11}(1650)$ (dashed lines). The corresponding kaon scattering angle is shown in each panel. Notation for experimental data is as in Fig.3.

Other promising observables for investigating the existence of narrow resonance at 1650 MeV are the differential cross section shown in Fig.6 and the beam-recoil double polarization $C_x$ shown in Fig.7. In Fig.6 we compare differential cross sections obtained from calculations without this resonance, along with the $P_{11}(1650)$ and $S_{11}(1650)$ resonances, with the available experimental data of elementary $K\Lambda$ photoproduction. It is clear from this figure that both $P_{11}(1650)$ and $S_{11}(1650)$ states could become good candidates for this narrow resonance. However, the two states produce a quite different effect on the differential cross section, i.e. whereas the $P_{11}(1650)$ resonance produces a peak at $W = 1650$ MeV, the $S_{11}(1650)$ state leads to a dip at this energy point. Since in the effect of the $P_{11}$ changes sign at $\theta_K^{c.m.} \approx 90°$, its contribution disappears in the total cross section after an integration over a complete angle. This is clearly in contrast to the effect of the $S_{11}$ state. Thus, it is obvious that experimental data, especially at backward angles, are able to pin down the correct state of this resonance.

Another possible observable is the beam-recoil double polarization shown in Fig.7. This polarization can be directly measured once we have polarized beam, since the polarization of recoiled $\Lambda$ can be determined directly from its decay. Comparison between the result of our calculation and experimental data is shown in Fig.7, where we can see that clear signal is produced by the two states. Since both states produce almost a similar effect, this observables apparently cannot be used to determine the correct resonance state as in the case of differential cross section. Nevertheless, measurement of this observable is still desired to eliminate the uncertainties in the resonance mass and total decay width. Furthermore, as seen in Fig.7 the accuracy of the presently available data is still unable to resolve the effect of this resonance. In view of this, it is important to improve the statistic of the future experiments.

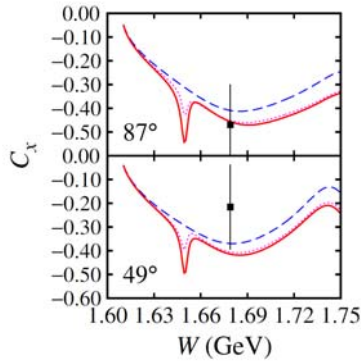

**FIGURE 7.** As in Fig.6, but for the beam-recoil double polarization $C_x$.

## SUMMARY AND CONCLUSION

We have investigated kaon photoproduction near its reaction threshold by using isobar models and found a structure in the recoiled hyperon polarization data which corresponds to a narrow resonance with a mass of 1650 MeV. The possible states of this resonance are $P_{11}(1650)$ and $S_{11}(1650)$. Since both states produce different effects in the experimental observables, future experimental measurements are able to resolve this issue. To this end, the most promising observables are the differential cross section at backward direction, recoiled hyperon polarization, and the beam-recoil double polarization. Such an experiment is apparently well suited for the Jefferson Lab as well as MAMI Mainz.

## ACKNOWLEDGMENTS

This work has been supported in part by the University of Indonesia and the Indonesian Ministry of Education and Culture through a Competence Grant scheme.